# PITFALLS IN GPR DATA INTERPRETATION: FALSE REFLECTORS DETECTED IN LUNAR RADAR CROSS SECTIONS BY CHANG'E-3

Chunlai Li, Shuguo Xing, Sebastian E. Lauro, Yan Su, Shun Dai, Jianqing Feng, Barbara Cosciotti, Federico Di Paolo, Elisabetta Mattei, Yuan Xiao, Chunyu Ding, Elena Pettinelli

*Abstract*—**Chang'e-3(CE-3) has been the first spacecraft to soft-land on the Moon since the Soviet Union's Luna 24 in 1976. The spacecraft arrived at Mare Imbrium on December 14, 2013 and the same day, Yutu lunar rover separated from lander to start its exploration of the surface and the subsurface around the landing site. The rover was equipped, among other instruments, with two Lunar Penetrating Radar systems (LPR) having a working frequency of 60 and 500 MHz. The radars acquired data for about two weeks while the rover was slowly moving along a path of about 114 m. At Navigation point N0209 the rover got stacked into the lunar soil and after that only data at fixed position could be collected. The low frequency radar data have been analyzed by different authors and published in two different papers, which reported totally controversial interpretations of the radar cross sections. The present study is devoted to resolve such controversy carefully analyzing and comparing the data collected on the Moon by Yutu rover and on Earth by a prototype of LRP mounted onboard a model of the CE-3 lunar rover. Such analysis demonstrates that the deep radar features previously ascribed to the lunar shallow stratigraphy are not real reflectors, rather they are signal artefacts probably generated by the system and its electromagnetic interaction with the metallic rover.**

## I. INTRODUCTION

Planetary subsurfaces are particularly suitable environments for Ground Penetrating Radar (GPR) investigations as they

This work was supported by the National Natural Science Foundation of China under Grant No. 41403054. (*Corresponding author: Yan Su; Shuguo Xing, Elena Pettinelli.*)

C.L.Li, S.G.Xing, Y.Su, S.Dai, J.Q. Feng, Y.Xiao and C.Y.Ding are with the Key Laboratory of Lunar and Deep Space Exploration, National Astronomical Observatories, Chinese Academy of Sciences, Beijing 100012, China (e-mail: suyan@nao.cas.cn, xingsg@nao.cas.cn).

S.E. Lauro, B.Cosciotti, F. Di Paolo, E.Mattei and E.Pettinelli are with the Mathematics and Physics Department of Roma Tre University, Roma 00146, Italy (e-mail: pettinelli@fis.uniroma3.it).

TABLE I
THE MAIN PLANETARY GROUND PENETRATING RADAR EQUIPMENTS

| Instrument | Mission (Year) | Object | Features |
|---|---|---|---|
| ALSE | Apollo 17 (1972) | Moon | Orbiter; 5-5.5MHz |
| LRS | KAGUYA (2007) | Moon | Orbiter; 4-6MHz 1MHz,15MHz Optinal |
| LPR | Chang'e-3 (2013) | Moon | Rover; 40MHz-80MHz; 250MHz-750MHz |
| CONSERT | Rosetta (2004) | 67P/C-G comet | Orbiter-Lander; 90 MHz |
| MARSIS | Mars Express (2003) | Mars | Orbiter; 1.3-2.3,2.5-3.5, 3.5-4.5,4.5-5.5MHz |
| SHARAD | MRO (2005) | Mars | Orbiter; 15MHz-25MHz |

are usually dry and cold, allowing good penetration and low attenuation of the radar signals. GPR is a well-established and mature technology for Earth applications [1], but is still in its infancy in planetary exploration. Indeed, since the early days of GPR development [2], only few space missions have been equipped with a subsurface radar instrument (see Table 1), even though the interest for this type of geophysical technique has grown quite rapidly in time. The role of orbiting subsurface radars have been fundamental in past [3] and still ongoing missions to the Moon and Mars [4],[5],[6] and will be essential to proof the existence of liquid water inside the icy crusts of the Jovian moons[7]. So far, among the successful missions reported in Table 1, Chang'e-3 (CE-3) Lunar Penetrating Radar (LPR) represents the first and only radar instrument employed onboard a rover [8], even though in the near future various rovers equipped with a GPR are expected to land on Mars [9],[10],[11] and on the Moon [12],[13].

The concept of radars mounted on a moving vehicle with the antennas operating near the surface is particularly appealing as they can provide high resolution electromagnetic imaging of the subsurface stratigraphy at different depth, depending on the antenna frequency employed. Such imaging could be used to choose the best location for drilling [11] or could be processed to extract quantitative information on the electromagnetic properties of the subsurface to better constrain orbiting radar data (ground truth) or geological interpretations [14].

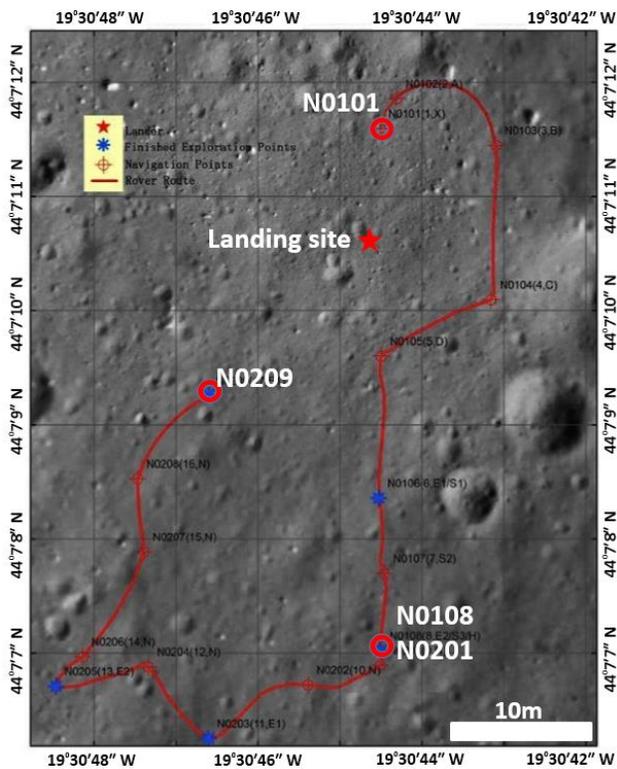

Fig. 1. The traverse path of the LPR on the moon. The red star represents the CE-3 landing site and the red line means the rover route. The background topographic picture was taken by the descending camera attached to the lander as it descended.

One of the major drawbacks in using near surface antennas is the clutter generated by objects on or above the ground, a problem well known in common applications of GPR on Earth and pointed out by several authors as early as the '90s [15]. In some circumstances, especially when the direction of the antenna dipoles is perpendicular to the objects [16], the radar cross sections can be affected by strong events coming from objects located above the survey line like metal objects, threes, boulders, walls, etc; these events are usually larger than those coming from subsurface reflectors because radar signal in the ground attenuate exponentially whereas signals propagating in the air attenuate geometrically [15]. Furthermore, as the dielectric contrast between air and surface objects is strong, even far distance targets can produce overlapping events that can be interpreted as real reflectors [17]. Particular attention should be paid at sub parallel events coming from above ground targets because they are difficult to be recognized ([16], [17]) and can lead to erroneous geological interpretation.

In planetary exploration the type of objects present above ground are limited to rocks, boulders and topographic reliefs, however the main problem is represented by the interaction between the emitted GPR signals and the metallic rover [18], [19]. The interference from the body of the rover can be quite strong if the antennas are unshielded and/or elevated above the ground and can introduce artefacts that can mask the subsurface events or "create" false reflectors in the radar cross section [19]. The main source of such artifacts is the energy reflected directly from the rover or after reflection from the surface which can originate ring-down periodic signals that can be interpreted as stratigraphy or multiple reflections [20]. Moreover, further sources of clutter could derive from spurious signals generated by electronic systems and connections, wheel motion or other instruments onboard the rover. Note that such artefacts could be filtered out applying different processing techniques (e.g., Solimene et al., 2014).

The above mentioned artefacts and clutter are well visible on LPR data, especially on those collected with the 60MHz dipolar antennas which were mounted above ground on the back of Yutu rover [21]. In fact the top part of the radar sections collected at this frequency on the Moon is systematically affected by a large ringing due to the antennas-rover coupling whereas, in the lower part the radar signal is quite weak and difficult to be interpreted. For this reason, so far, most of the work on LRP radar data have been focused on the 500MHz [22-25] and only two articles attempted to interpret some deep reflectors detectable on the 60MHz data as stratigraphic interfaces, obtaining totally controversial results [26], [27]. The present paper is devoted to resolve such controversy by a careful analysis and comparison of data collected on the Moon by Yutu rover and on Earth by a prototype of LRP mounted onboard a model of the CE-3 lunar rover [28]. Our analysis shows that the deep reflectors cannot be ascribed to real geological interfaces as clamed in previous works [26], [27], rather they are artefacts probably introduced by the overall system. The paper is organized as follows: In section Ⅱ, the LPR and moon operations are presented. In Section Ⅲ, LPR calibration on earth and previous studies are listed. Based on radar equation, the reflector detectability of LPR are calculated, which is aimed to verify whether the radar can reach those depths, which is shown in Section Ⅳ. Two detailed data analysis and effect of noise on reflectors detectability and Time-Frequency analysis applying S transform are conducted in Section Ⅴ. Section Ⅵ is the discussion. Finally, the conclusions are drawn in Section Ⅶ.

## Ⅱ. LPR AND MOON OPERATIONS

CE-3 mission is part of the second phase of China Lunar Exploration Project, which is aimed at exploring the Moon with rovers and landers, and follows the first phase performed by Chang'e-1(CE-1) and Chang'e-2(CE-2) orbiting spacecrafts [29]. CE-3 mission was successfully launched from Xi-Chang satellite launching center on December 2nd, 2013 and landed on the northern Mare Imbrium after 12 days. It was the first spacecraft to soft-land on the Moon since the Soviet Union's Luna 24 in 1976. The same day of landing (December 14, 2013 at 20.35 UTC), Yutu lunar rover separated from lander and started its march to explore the surface and the subsurface around the landing site. Yutu payload included an active particle induced X-ray spectrometer (APXS), a visible to near-infrared (450–945 nm) imaging spectrometer and short-wave infrared (900–2,395 nm) spectrometer (VNIS), and the LPR, together with a stereo camera and a navigating camera. The main goal of the radar was the estimation of the thickness of lunar regolith and the detection of the lunar subsurface structure along the traverse path. LPR is an ultra-broadband radar operating in time domain and equipped with two sets of antennas: i) a pair of low frequency monopole antennas (1150 mm long 12 mm diameter and separated by about 80 cm) with 60 MHz center frequency and 40 MHz bandwidth (i.e., 40-80 MHz), suspended 60 cm above the ground on the back of the rover (see Fig.1); ii) a set of one transmitter and two receiver bowtie antennas (336 mm long and 120 mm wide) operating at 500 MHz center frequency, with 500 MHz bandwidth (i.e., 250-750 MHz). These antennas were located at the bottom of lunar rover, about 30 cm above the ground, and separated 160 mm from each other [8]. We refer to these sets as channel one and channel two, respectively. Channel one has a system gain of 152 dB and channel two of 133 dB. Detailed description of the radar system can be found in [8].

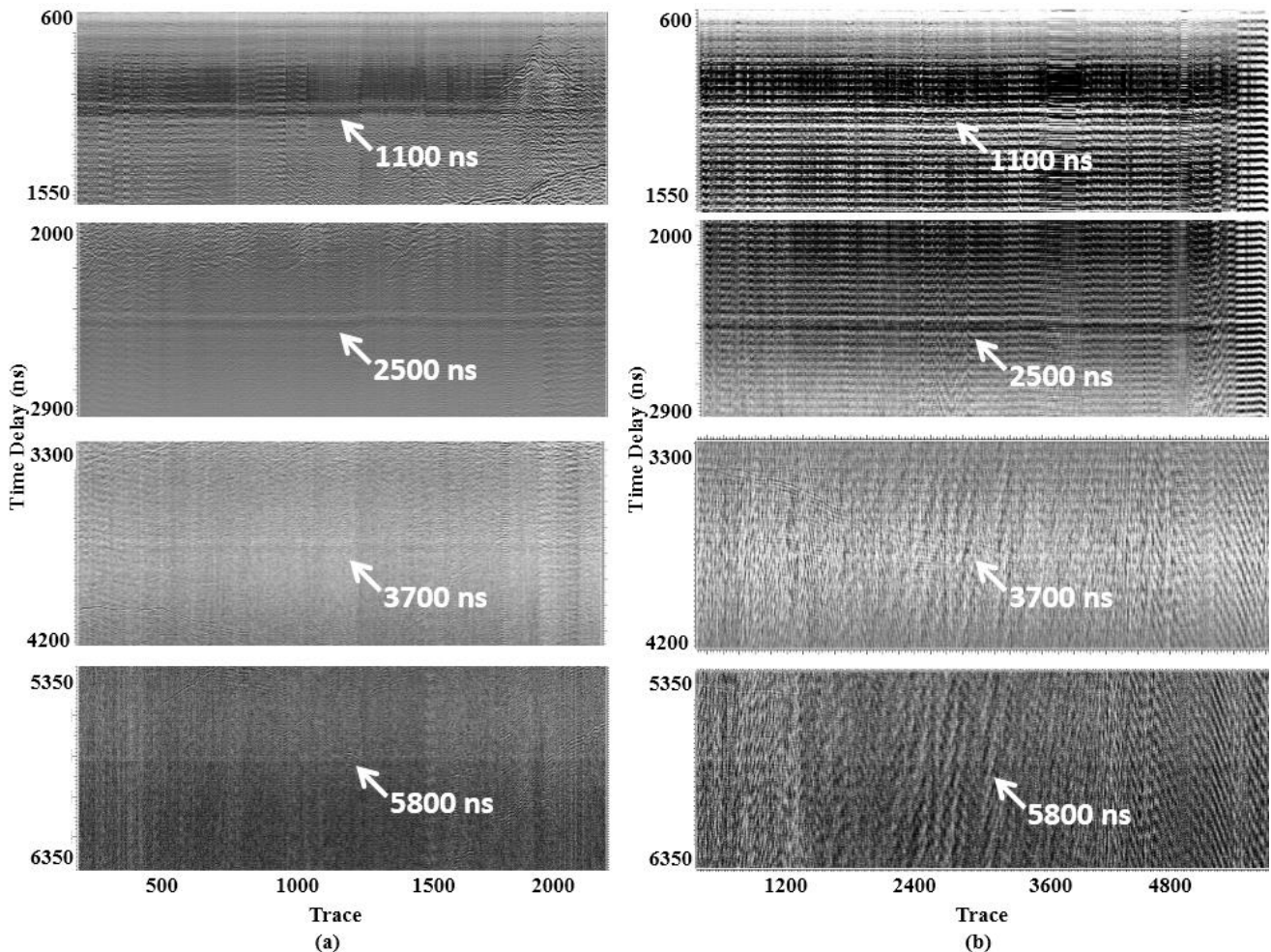

Fig. 2. Deep radar features detected on glacier (a) and loess deposit (b). Note that the reflectors are located at the same time depth (indicated by the white arrow) but they are better visible on the glacier data as noise and ringing are lower, especially at larger depth, with respect to loess deposit data.

The LPR operational phases on the Moon can be divided into three stages (see Fig. 2). In the first stage the rover traveled 56 m in 10 days, moving from navigation point N0101 to point N0108. During this phase several radar parameters (e.g., system gain, time window and attenuation settings) were tested to determine the performance of the system for the subsequent data acquisition. The second stage started at point N0201 and after 58 m reached point N0209; in this phase the radar acquired data from both channels using the acquisition parameters previously set. Unfortunately, after this phase the rover got stack in the Moon soil and could not move any further. As a consequence in the last phase the radar acquired data only at a fix position (i.e., at point N0209).

## III. LPR Calibration on Earth and Previous Studies

LPR system was calibrated and tested on Earth before launching, using an LPR prototype with a 1:1 simplified model of the CE-3 lunar rover, equipped with a GPS to track the radar profile location. Such prototype reproduce quite well the performance of the system even though, in the actual radar, specific isolators and filters to avoid the influence of communication signals between the lunar rover and the lunar lander were also installed [28]. The testing campaign was carried out by the Ground Research and Application System, NAOC and Institute of the Electronics of Chinese Academy of Sciences on three different types of ground: a glacier, a loess deposit and an artificial lunar soil [28]. In the first two sites both antennas (60 MHz and 500 MHz) were tested, whereas in the artificial lunar soil, being as shallow as 15-70 cm, only the 500 MHz was used. However, as in this paper we are interested in the 60 MHz data, we would not describe or comment any further the data collected with the higher frequency antennas. The best performance in terms of maximum penetration depth of the LPR prototype was achieved on the glacier experiment where, assuming a permittivity of 3.2, geological structures as deep as 152 m were recognized in the radar cross sections [28]. However, the data collected with the 60 MHz antennas on glacier also show, at various time depth (i.e., 1100 ns, 2500 ns, 3700 ns and 5800 ns) weak continues signals which appear on the radar cross sections as quasi-horizontal features of constant amplitude (see Fig.2a). In particular, the shallowest one (1100 ns) seems to cut some geological structures visible on the right side of the section whereas the others seem to only superimpose to the noise. The same radar features are present at similar time depths in the sections collected on loess deposit (see Fig.2b) even though, in this case, the deepest interfaces (3700 ns and 5800 ns) are difficult to be recognized as they are almost totally buried in the noise. Nevertheless, none of the papers published on LPR calibration and testing have mentioned or discussed the presence and origin of such features.

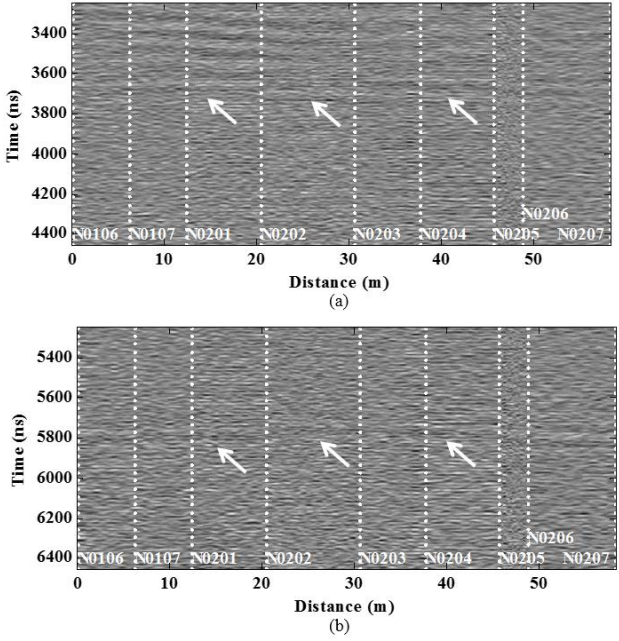

Fig. 3. Radar features on Moon data collected at 60 MHz from navigation point N0106 to point N0207. The vertical dashed lines indicate different navigation points which refer also to different days of acquisition. White arrows indicate two deep hypothetical reflectors.

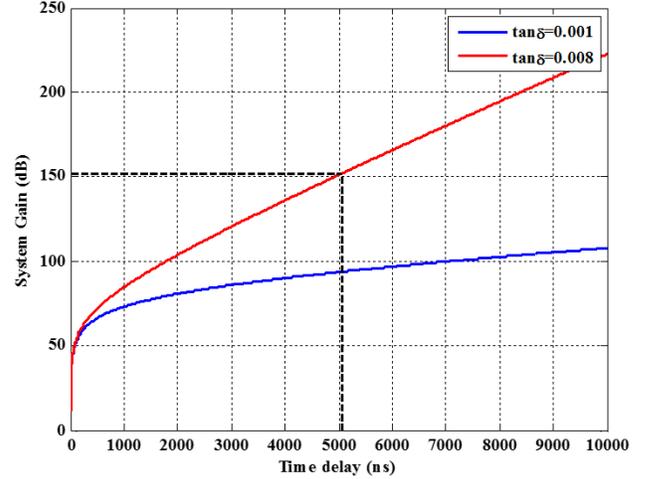

Fig. 4. Relationship between system gain and time delay computed using Equation (2). Blue line indicates the most favorable case for signal penetration (low loss), red line the worst case.

Conversely, a detailed discussion on the two deepest features (3700 ns and 5800 ns), which are also visible on the radar sections collected on the Moon (see Fig.3), is present in two different papers [26],[27]. Indeed both studies assume that such features are real reflectors generated by some geological discontinuity. In particular, Xiao and co-workers interpreted such reflectors as two layers representing different episodes of lava eruptions [26], whereas Zhang and co-authors interpreted the reflectors as buried regolith layers between different basalt units [27]. Note that in the data collected on the Moon the two shallow radar features (at 1100 ns and 2500 ns) are not visible as totally buried in the strong ringing present in the first 3000 ns of the time window (cf. Fig. 5 in [27]).

A careful comparison between Earth and Moon data like, for example, that illustrated in Fig.2 and Fig.3, seems to unambiguously indicate that such features are not real reflectors. In fact, it is highly improbable to find the same features in such different survey environments and operating conditions. Nevertheless a simple qualitative analysis cannot be considered a robust and definite evidence as, accounting also for the measurement uncertainties, some fortuitous coincidence in time depth cannot be excluded. Therefore, in the following we will perform a detailed quantitative analysis to prove that such features do not correspond to real geological lunar structures but rather they are generated by the electronics and/or by the electromagnetic interaction between the system and the rover.

## IV. RADAR EQUATION AND SIGNAL PENETRATION

As a first step, we have tested the maximum signal penetration depth in lunar soil using the radar range equation and the parameters of the LPR at 60 MHz. Given the equation [30]

$$P_r = P_T \frac{G_{Tx} G_{Rx} \lambda^2 \sigma}{(4\pi)^3 D^4} e^{\left(\frac{-2\omega \tan \delta D}{\nu}\right)} \quad (1)$$

Where:

$P_T$ is the transmitting power;

$P_r$ is the receiving power;

$G_{Rx}$ is the transmitting antenna gain;

$G_{Tx}$ is the receiving antenna gain;

$\lambda$ is the wavelength in the medium;
$\nu$ is the velocity in the medium;

| Model | σ(a.u.) |
|---|---|
| High-voltage-on | 2.17±0.16 |
| High-voltage-off | 2.16±0.16 |

$D$ is the depth of layer, $D = v \times t / 2$; $t$ is the two way travel time.

$\tan\delta$ is loss tangent;

$\sigma$ is radar cross section;

$\omega = 2\pi f$; $f = 60 MHz$;

We can re-arrange the terms to estimate the system gain $G_{sys}$ as:

$$G_{sys} = \frac{P_T G_{Rx} G_{Tx}}{P_{min}} = \frac{(4\pi)^3 D^4}{\lambda^2 \sigma} e^{\frac{2\omega \tan\delta D}{v}} \qquad (2)$$

Where $P_{min}$ is the minimum detectable power at the receiver. The parameters in Equation (2) are taken from Table 1 in[28] and the depth D is computed on the basis of the velocity v=0.11 m/ns which is an average value for Mare Imbrium according to [6]. Furthermore we have assumed $\sigma = \pi D^2$, which represents the maximum back scattering (total reflection) due to a flat and smooth interface. The choice of the value of tanδ is quite critical because it is not well constrained, therefore we can only consider a wide range of values, that is $10^{-3}$ to $8 \times 10^{-3}$, as suggested by [6]. Fig. 4 illustrates the theoretical trend of the system gain values vs. time depth for the above mentioned loss tangent range boundaries. The blue line refers to the lower loss tangent (tanδ=10$^{-3}$) and the red line to the highest value (8x10$^{-3}$). It is evident that in this interval of values the dynamic range of channel one (152dB) is suitable to detect bright reflectors up to a depth of about 5000ns.

## V. DATA ANALYSIS AND RESULTS

### A. Effect of Noise on reflectors detectability

As a second step, we have evaluated the reflector detectability on Moon data estimating the noise level and comparing such level with the signal amplitude of the deep reflectors. To this aim we analyzed two datasets: the first one collected with the transmitter switched off (High-Voltage-Off mode) at navigation point N0209 (from 2014-02-15T22:51:38 to 2014-02-15T23:56:34), and the other with the transmitter on (High-Voltage-On mode) acquired from navigation point N0101 to N0208. The estimation was made considering a subset of 150 traces extracted at N0201 (High-Voltage-On) and N0209 (High-Voltage-Off) respectively and computing the noise level in terms of standard deviation of the signal amplitude. In High-Voltage-Off mode the noise was calculated on the total time window (0-10000 ns) whereas in the High-Voltage-On mode only the bottom part of the time window (i.e., 6000 - 10000 ns) was considered to minimize the effects of weak reflectors present above such time depth. As expected, the values computed for the two modes are very similar (see Table II) so we have assumed for the noise level a standard deviation σ=2.16. In Fig.5 it is reported the comparison between the amplitude values of all traces collected in the time windows 3600 - 3800 ns and 5700 - 5900 ns (i.e., the time intervals where the hypothetical reflectors have been detected) with the noise level computed at 1σ, 2σ and 3σ. From the figure it is clear that the signal amplitude level in the two time intervals is quite similar and it barely exceeds the 3σ noise level.

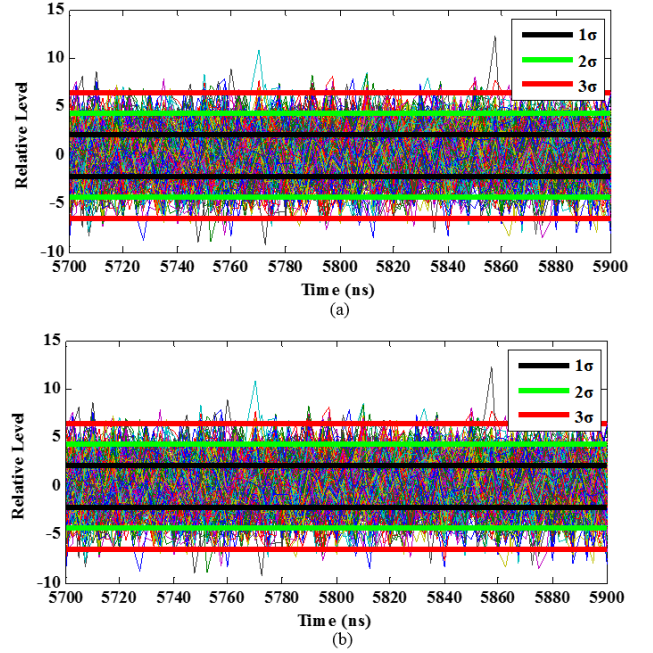

Fig. 5. Comparison between signal amplitude and noise level for the Moon data: (a) The hypothetical reflector between 3600 and 3800 ns; (b) The hypothetical reflector between 5700 and 5900 ns.

### B. Time-Frequency analysis applying S-transform

Further information about the nature of the radar features under investigation can be searched looking at their spectral content using different time-frequency representations (like, for example, Short Time Fourier Transform, Wigner-Wille distribution or S-transform) [31]. In this work we have chosen to use the S-transform as it represents a good compromise between frequency-time resolution and simple spectrum interpretation. This technique is conceptually similar to the continuous wavelet transform (CWT) and it is based on a moving and scalable localizing Gaussian window. The S-transform S(t,f) of the signal x(t) is defined as:

$$S(t,f) = \int x(\tau) \frac{|f|}{\sqrt{2\pi}} e^{\frac{-f^2(t-\tau)^2}{2}} e^{-i2\pi f\tau} d\tau \qquad (3)$$

where f is the frequency.

First of all we tested such technique on the data collected on the glacier, analyzing the signals reflected by natural discontinuities (i.e., the dipping reflector detected by [28]) present in the ice layer at a time depth between 1400 and 1800 ns. Fig. 6 shows the results of the S-transform analysis applied to the entire time window of a single trace. More specifically, panel 6(a) illustrates the time trace and panel 6(b) the time-frequency trace. In the latter, the insert box shows the frequency content of the reflector at about 1600 ns which exhibits a spectrum centered at 60 MHz in full agreement with the central frequency of channel one transmitting antenna. We then applied the same procedure to the complete glacier data set (Fig. 7a) and we extracted from the S-transform only the 40-80MHz components in order to generate the time-amplitude image reported in Fig. 7b. The comparison between these images shows that the natural discontinuity inside the glacier produce echoes that preserve the original frequency content emitted by the antenna.

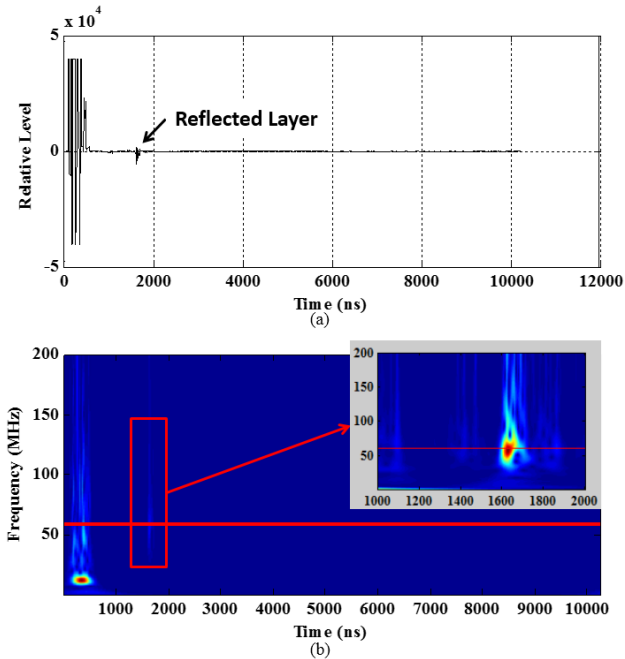

Fig. 6. An example of the S-transform based on glacier data. (a) One trace of radar data. (b) S-Transform of (a).

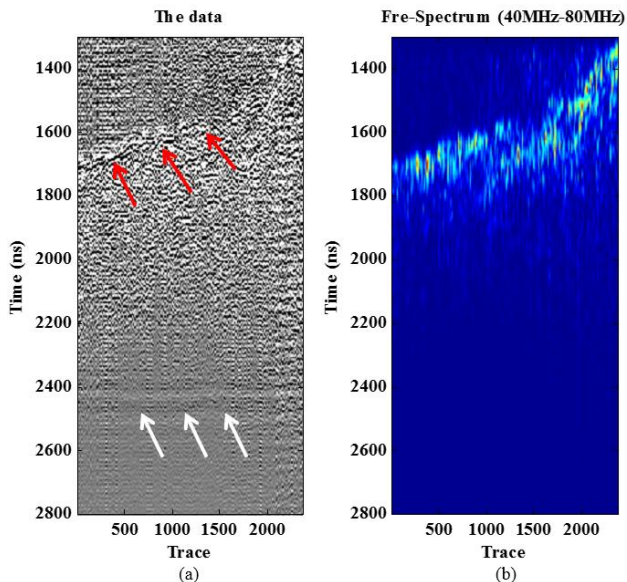

Fig. 7. The S-transform of the Glacier data (a) The chosen part (b) The S-transform result with frequency spectrum (40MHz-80MHz)

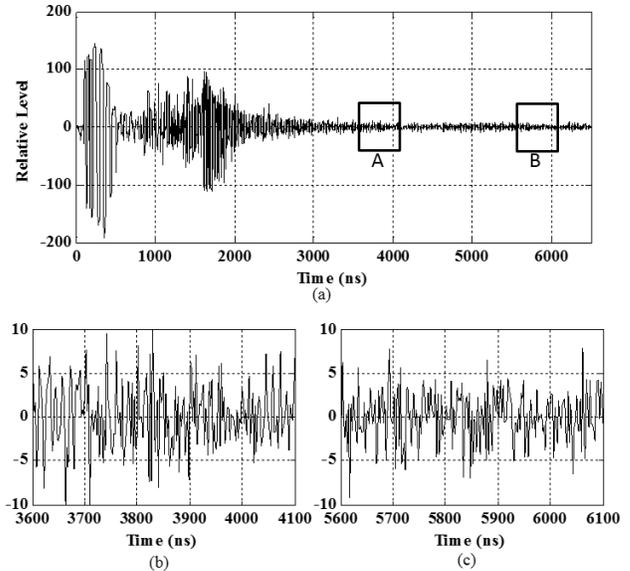

Fig. 8. Example of radar trace collected on the glacier. (a) Boxes labelled A and B indicate the position of the deepest events also present on Moon data. (b) Enlarged view of data in box A; (c) Enlarged view of data in box B.

As a last test on the glacier data, we have selected the time interval around the four hypothetical reflectors (see section Ⅲ ) to study the frequency content of such features. However, because the amplitude of these features is of the same order of magnitude of the noise level (see Fig. 8), as also noticed for the Moon data (cf. Section Ⅲ), the S-transform has been applied on a single stacked trace averaging 1000 traces. Fig. 8 illustrates, as an example, the level of the signal on a radar trace in the time interval around the position of the two deepest hypothetical reflectors whereas Fig. 9 shows the stacked trace (panel a), the signal in the time intervals of interest (panel b) and the corresponding S-transform (panel c). Note that in each graph of panel (b) the red line is a sinusoidal signal having time position and frequency given by the coordinates (f, t) of the maximum value of the S-transforms illustrated in panel (c). The spectrum central frequency of all four hypothetical reflectors is substantially the same (about 12 MHz), and differently from the previous case (cf. Fig. 6 and Fig. 7), it is well below the bandwidth of the transmitted signal (40-80 MHz).

Finally, the same procedure described above was applied to the Moon data, computing the average trace from 1000 traces collected at navigation point N0201. Fig. 10 illustrates the results of the analysis in a similar fashion as Fig. 9 but only for the two deepest hypothetical reflectors as the shallow ones cannot be extracted from the ringing (see Section Ⅲ). The results are in good agreement with those obtained for the glacier as the time depth and the spectrum central frequency (about 12 MHz) values of the features are quite similar.

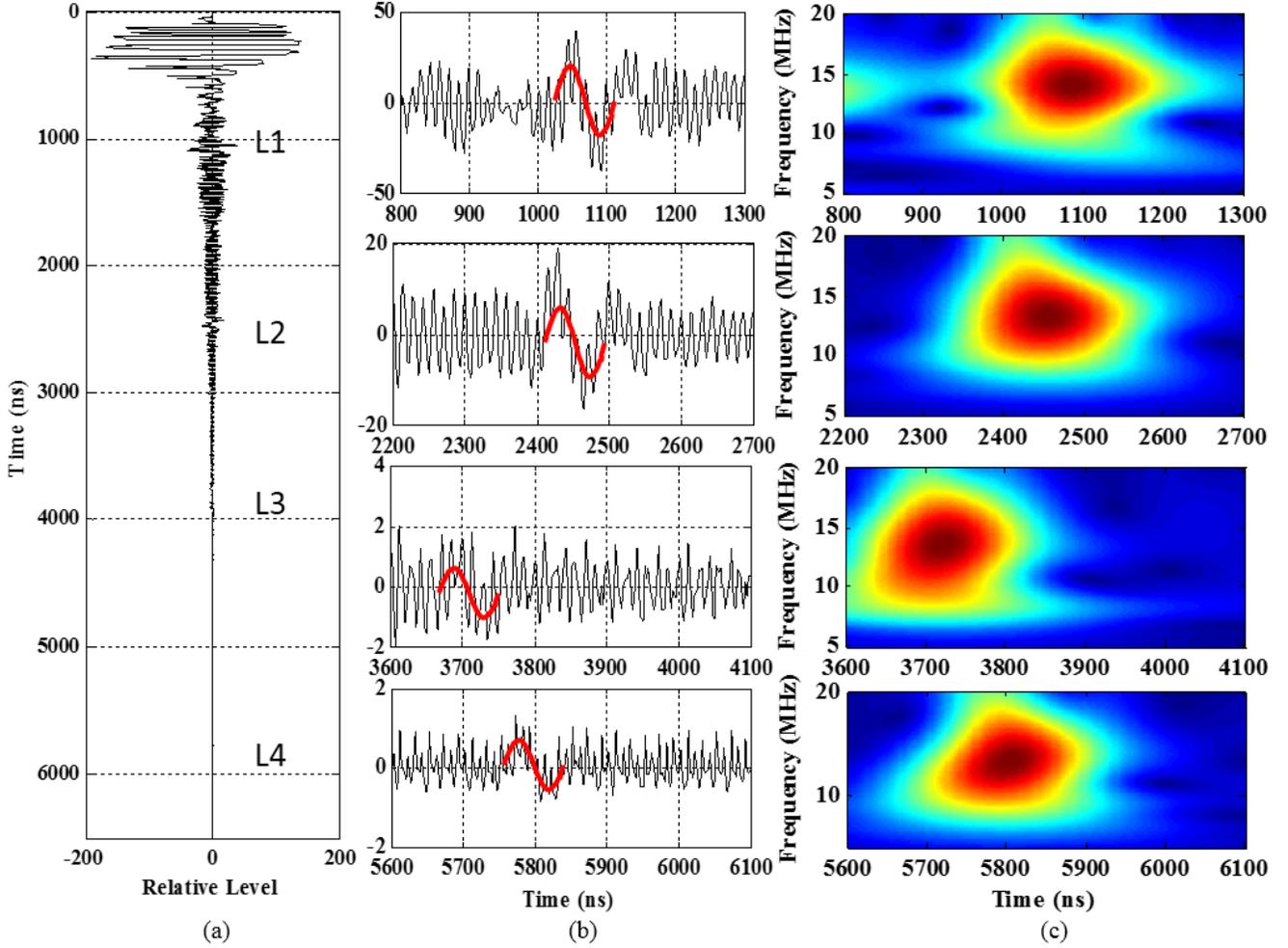

Fig. 9. Results of the S-transform applied to the averaged trace computed from glacier data. (a) Position of the time intervals where the hypothetical reflectors are located. (b) Enlarged view of the time intervals and (c) Relevant S-transform images.

## VI. Discussion

As highlighted in Section Ⅲ (see Fig. 2 and Fig. 3) shape, amplitude and position of the hypothetical reflectors appear to indicate that they are stationary disturbances always present in the radar cross sections and well visible only if some specific processing is applied to enhance them above the noise level (see, for example, Supporting information of [27]). In particular, a careful comparison between Fig. 2 and Fig. 3 highlights that the shape of such features on the Moon data are slightly curved if compared with those present, at the same time depth, in the glacier and loess data. This fact is probably due to the difference in terms of acquisition time of the radar data set; indeed, the data collected on the glacier and loess terrains was made in a single day, whereas the data collected on the Moon refers to fifteen different days and thus different operating modes of the rover and of the other instruments. As a consequence, the lunar radar cross section is an assemblage of different data sets that do not properly align, even though inside single blocks of data (i.e., between two subsequent navigation points) the radar features are essentially horizontal (see also Fig. 5 in [27]).

The theoretical computation performed assuming an ideal reflector and considering the expected values for the loss tangent in Mare Imbrium have shown that it cannot be excluded that the LPR dynamic range (152 dB) could be sufficient to detect reflectors as deep as 300 m. However, the quantitative analysis performed on the signal amplitude support the hypothesis that the deep radar features under question do not come from real geological structures. The analysis performed on the noise highlights that there is no significant difference in terms of amplitude level between the noise and the signals in the time windows associated to the two hypothetical reflectors. In fact, the signal amplitude level for the lunar deep reflectors is of the order of $3\sigma$, however similar values can also be found for the data collected on the glacier. Furthermore, the similarity between the amplitude levels of the signals coming from different depths poses serious questions about the reliability of the data. Indeed, if we assume that the signals coming from the first deep reflector (at about 3700 ns) are real and their small amplitude is due to the propagation in the soil (non-negligible attenuation), it is physically unlikely that the second reflector (at about 5800 ns) could maintain the same amplitude after over 2000 ns of propagation.

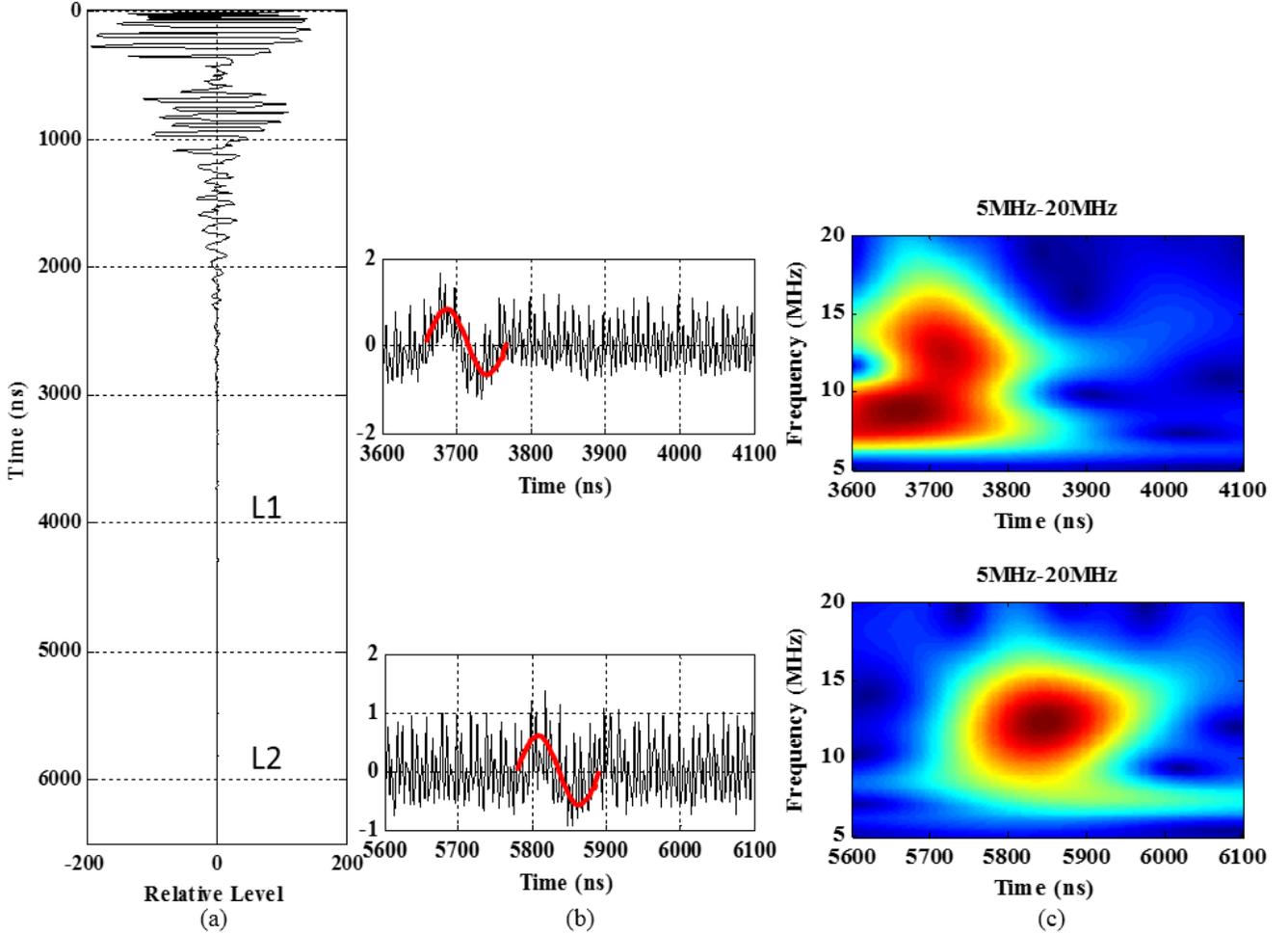

Fig. 10. Results of the S-transform applied to the averaged trace computed from moon data. (a) Position of the time intervals where the hypothetical reflectors are located; (b) Enlarged view of the time intervals and (c) Relevant S-transform images.

The S-transform analysis adds another important piece of information to clarify the nature of radar features interpreted by various authors ([26], [27]) as real reflectors. The main outcome of such analysis is the presence of a systematic frequency shift in the radar signals regardless survey location (glacier or lunar terrain) and time delay. Indeed, for both terrestrial and lunar radar data, the values of the central frequency (10-15 MHz) of the signals hypothetically coming from deep interfaces are well below the central frequency of the transmitted signals (60 MHz). On Earth, it is well known that a radar signal can experience an apparent shift in frequency when it propagates trough a soil acting as low-pass filter[32],[33]; such shift is ruled by the relationship between attenuation and frequency. In the case of signals propagating through the ice (glacier survey), in which the attenuation is frequency independent, no frequency shift should be expected and any signal coming from a real reflector should preserve the original frequency content. The analysis of the signals coming from the interface located between 1400-1800 ns in the glacier radar cross section are in full agreement with this statement (see Fig.6). On the contrary, the signal coming from the hypothetical deep reflectors have a frequency content which cannot be justified by the propagation in the subsurface.

Differently from ice, the materials composing Mare Imbrium terrains can act as low-pass filter and can produce a shift in frequency as, according to our knowledge about lunar materials, the radar attenuation is frequency dependent[34],[35]. However, such shift can be appreciated only if the attenuation (or loss tangent) in the material and the time depth of the reflector are large enough. A way to verify if the shift could be real is to compute the value of the loss tangent (tanδ) that would have produced such effect. In practice, considering the two hypothetical reflectors (at about 3700 ns and 5800 ns), tanδ can be estimated applying Equation (11) in [35] and assuming a bandwidth of 40 MHz. For the shallower reflector we found $\tan \delta_{3700ns} = 1.6 \times 10^{-2}$ and for the deepest one $\tan \delta_{5800ns} = 0.8 \times 10^{-2}$; both values are larger than those expected for Mare Imbrium terrains (see Fig. 4). This result poses again a question about the reliability of the radar data. In fact, if we assume that the computed loss values were real, the maximum penetration depth achievable by the system on the Moon would be lower than 3700 ns (see Section IV) and no deep signal could be actually detected.

Finally, despite a total disagreement about the interpretation of the radar features between the present study and [27], our spectral analysis results are strongly supported by the work of Zhang and co-workers [27]. In fact, these authors found that the only way to extract from the noise the signals associated to the hypothetical deep reflectors is to apply a band-pass filter between 4 and 30 MHz that is, outside the frequency band of the transmitted signal.

## VII. Conclusion

In the present study we have analyzed the characteristics of the radar signals collected on the Moon by LPR using channel one antennas. The scope of the work was to clarify the origin of specific radar features visible in the radar cross sections and previously interpreted as real reflectors associated to the layering structure of the lunar subsoil. Such analysis was based on three different approaches: i) a qualitative comparison between the radar data collected on the Earth and on the Moon using channel one LPR system; ii) the amplitude of the radar signal vs. the background noise level; and iii) the comparison between the signal frequency content of the transmitted and hypothetically reflected signals. We found that the overall results are robust, fully consistent and totally unambiguous. Therefore we can conclude that the deep radar features are not real reflectors rather they are signal artefacts superimposed to the radar traces. Regarding the origin of these artefacts at the moment we can only speculate. Our study has shown that these signals are present, with almost identical characteristics, in both terrestrial and lunar data, suggesting that they are probably generated by the electronic of the system and/or the radar-rover coupling. In 2018 a new opportunity to study the Moon subsurface will be offered by CE-4 mission, which will be equipped with the same radar as CE-3. It would be of paramount importance to fully understand the nature of such strong disturbance and, possibly, eliminate it to obtain a reliable view of the lunar shallow geology.

### Acknowledgment

The data are provided by the Lunar and Deep Space Exploration Department, National Astronomical Observatories, Chinese Academy of Sciences (NAOC) and available at http://moon.bao.ac.cn.